\documentclass[sigchi-a, authorversion]{acmart}
\usepackage{booktabs} 
\usepackage{ccicons}  

\setcopyright{acmcopyright}
\acmDOI{10.475/123_4}

\acmISBN{123-4567-24-567/08/06}

\acmConference[CHI 2019 Workshop on Speech Interface Interactions]{Workshop on Mapping Theoretical and Methodological Perspectives for Understanding Speech Interface Interactions}{4-9 May 2019}{Glasgow, UK}
\acmYear{2019}
\copyrightyear{2019}

\acmPrice{15.00}

\begin{document}
\title{A `Canny' Approach to Spoken Language Interfaces}

\author{Roger K. Moore}
\affiliation{%
  \institution{University of Sheffield}
  \city{Sheffield}
  \state{S. Yorks.}
  \postcode{S1 4DP}
  \country{UK} }
\email{r.k.moore@sheffield.ac.uk}

\renewcommand{\shortauthors}{R. K. Moore}

\begin{abstract}  
Voice-enabled artefacts such as Amazon Echo are very popular, but there appears to be a `habitability gap' whereby users fail to engage with the full capabilities of the device.  This position paper draws a parallel with the `uncanny valley' effect, thereby proposing a solution based on aligning the visual, vocal, behavioural and cognitive affordances of future voice-enabled devices.
\end{abstract}

\keywords{voice enabled devices; habitability gap; uncanny valley effect; aligned affordances}

\maketitle

\begin{marginfigure}
    \includegraphics[width=\marginparwidth]{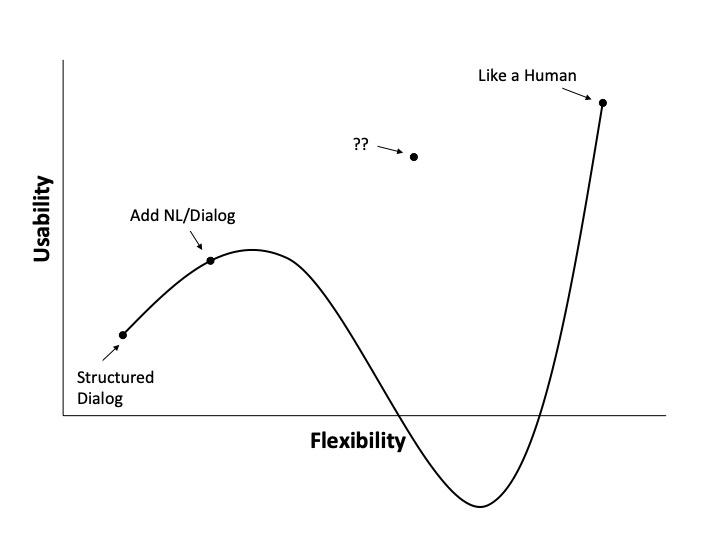}\Description{????}
    \caption{Increasing the flexibility of spoken language dialogue systems can lead to a `habitability gap' \cite{Philips2006}.}
    \label{fig:MP}
\end{marginfigure}

\begin{marginfigure}
    \includegraphics[width=\marginparwidth]{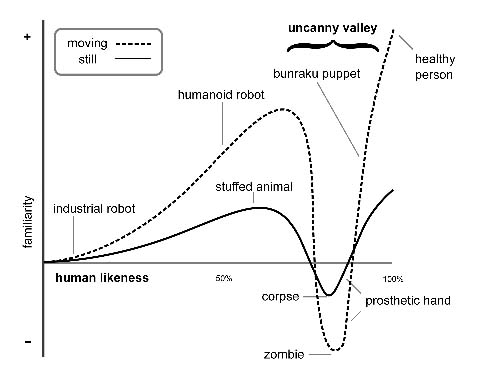}\Description{????}
    \caption{The `uncanny valley' effect \cite{Mori1970}.}
    \label{fig:UV}
\end{marginfigure}

\section{Introduction}

Recent years have witnessed astonishing progress in the development of voice-enabled artefacts such as \emph{Siri} (released by Apple in 2011) and \emph{Alexa} (released by Amazon in 2014).  For example, tens of millions of Alexa-enabled devices were sold worldwide over the 2017 Christmas holiday season, and Apple’s Siri had 41.4 million monthly active users in the U.S. as of July 2017 \cite{Boyd2018}.  Indeed, the appearance of such `intelligent' personal assistants is often hailed as a significant step along the road towards more natural interaction between human beings and future `autonomous social agents' (such as robots).

However, studies into the usage of such technology suggest that, far from engaging in a promised natural `conversational' interaction, users tend to resort to formulaic language and focus on a handful of niche applications which work for them \cite{Moore2016}.  Given the pace of technological development, it might be expected that the capabilities of such devices will improve steadily, but evidence suggests that there is a `habitability gap' (see Figure~\ref{fig:MP}) in which usability drops as flexibility increases \cite{Philips2006}.

\section{The Problem}

\begin{marginfigure}
    \includegraphics[width=\marginparwidth]{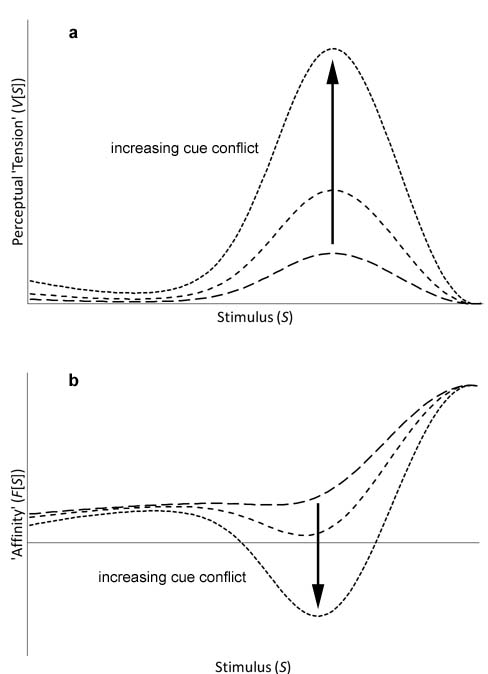}\Description{????}
    \caption{(a) Perceptual `tension' increases at a category boundary as a result of misaligned perceptual cues.  (b) Peaks in perceptual tension give rise to dips in `affinity', i.e. increases in ‘uncannyness’ \cite{Moore2012}.}
    \label{fig:PT}
\end{marginfigure}

It has been hypothesised that the habitability gap is a manifestation of the `uncanny valley' effect (see Figure~\ref{fig:UV}) whereby a near human-looking artefact (such as a humanoid robot) can trigger feelings of eeriness and repulsion \cite{Mori1970}.  In particular, a Bayesian model of the uncanny valley effect \cite{Moore2012} reveals that it can be caused by \emph{misaligned} perceptual cues (see Figure~\ref{fig:PT}).  Hence, a device with an \emph{inappropriate} voice can create unneccesary confusion in a user.  For example, the use of human-like voices for artificial devices encourages users to overestimate their linguistic and cognitive capabilities.

\section{A Way Forward}

The Bayesian model of the uncanny valley effect suggests that the habitability gap can only be avoided if the visual, vocal, behavioural and cognitive \emph{affordances} of an artefact are aligned.  Given that the state-of-the-art in these areas varies significantly, this means that the capabilities of an artificial agent should be determined by the affordance with the lowest capability.  In other words, emulating a human is a recipe for failure, rather ``\emph{it is better to be a good machine than a bad person}'' \cite{Balentine2007}.

So, the theoretical perspective discussed above suggests a whole-system design approach in which the characteristics of each element must be selected in accordance with the characteristics of all other elements.  In particular, the voice of an artefact should be selected by taking into account all other aspects of the design.  For example, the vocal tract length should be based on the physical size of the device, the vocal timbre should be based on the construction material, and the linguistic complexity of its utterances should be conditioned on its underlying cognitive abilities.  Failure to follow this approach will lead to the creation of yet more voice-enabled \emph{chimeras} sitting solidly in the habitability gap.

\section{Conclusion}

Notwithstanding the immense progress that has been made in voice-enabled artefacts, future progress depends on designers taking a whole-system perspective and ensuring that the visual, vocal, behavioural and cognitive affordances are aligned.  Inspiration can be taken from fictional characters in cinema and television \cite{Wilson2017}, and such an approach will open up a plethora of imaginative and yet appropriate voices \cite{Moore2017} - altogether a more `canny' approach to the development of intelligent communicative machines \cite{Moore2015}.

\newpage

\bibliography{RKM-position-paper}
\bibliographystyle{ACM-Reference-Format}

\end{document}